\documentclass[sigconf]{acmart}

\AtBeginDocument{%
  }
    
\usepackage{enumitem}
\usepackage{subcaption}
\usepackage{stfloats}
\usepackage{float}
\usepackage{array}

\usepackage{graphicx}
\usepackage{pifont}
\newcommand{\cmark}{\ding{51}}
\newcommand{\xmark}{\ding{55}}

\usepackage{xspace}
\newcommand{\vogue}{\textsc{Vogue}\xspace}

\copyrightyear{2026}
\acmYear{2026}
\setcopyright{cc}
\setcctype{by}
\acmConference[UMAP '26]{34th ACM Conference on User Modeling, Adaptation and Personalization}{June 08--11, 2026}{Gothenburg, Sweden}
\acmBooktitle{34th ACM Conference on User Modeling, Adaptation and Personalization (UMAP '26), June 08--11, 2026, Gothenburg, Sweden}
\acmDOI{10.1145/3774935.3806177}
\acmISBN{979-8-4007-2311-7/2026/06}

\begin{document}

\title{VOGUE: A Multimodal Dataset for Conversational Recommendation in Fashion}

\author{David Guo}
\authornote{These three authors contributed equally to this work.}
\affiliation{%
  \institution{University of Toronto}
  \city{Toronto}
  \state{Ontario}
  \country{Canada}}
\email{davidmy.guo@mail.utoronto.ca}

\author{Minqi Sun}
\authornotemark[1]
\affiliation{%
  \institution{University of Waterloo}
  \city{Waterloo}
  \state{Ontario}
  \country{Canada}}
\email{maggie.sun@uwaterloo.ca}

\author{Yilun Jiang}
\authornotemark[1]
\affiliation{%
  \institution{University of Waterloo}
  \city{Waterloo}
  \state{Ontario}
  \country{Canada}}
\email{yilun.jiang@uwaterloo.ca}

\author{Jiazhou Liang}
\affiliation{%
  \institution{University of Toronto}
  \city{Toronto}
  \state{Ontario}
  \country{Canada}}
\email{joe.liang@mail.utoronto.ca}

\author{Scott Sanner}
\affiliation{%
  \institution{University of Toronto}
  \city{Toronto}
  \state{Ontario}
  \country{Canada}}
\email{ssanner@mie.utoronto.ca}
\renewcommand{\shortauthors}{Guo, Sun, Jiang, et al.}



\begin{abstract}
Multimodal conversational recommendation has recently emerged as a promising paradigm for delivering personalized experiences through natural dialogue enriched by visual and contextual grounding. Yet, currently available multimodal conversational recommendation datasets remain limited: existing resources either simulate conversations, omit user history, or fail to collect sufficiently detailed feedback, all of which constrain the types of research and evaluation they support. 

To address these gaps, we introduce \vogue, a novel dataset of 60 human-human dialogues 
containing 2,100 granularly labeled utterances 
in realistic fashion shopping scenarios. Each dialogue is paired with a shared visual catalogue, item metadata, user fashion profiles/histories, and post-conversation ratings from both users (Seekers) and recommenders (Assistants). This design enables rigorous evaluation of conversational inference, including not only alignment between predicted and ground-truth preferences, but also calibration against full rating distributions and comparison with explicit and implicit user satisfaction signals. 

Our initial analyses of \vogue reveal distinctive dynamics of visually grounded dialogue, e.g., recommenders frequently recommend items simultaneously in 
feature-based groups, which creates 
distinct conversational phases bridged by Seeker critiques and refinements. Benchmarking Multimodal
Large Language Models against human Recommenders shows that while MLLMs approach human-level alignment in aggregate, they exhibit systematic distribution errors in reproducing human ratings and struggle to generalize preference inference beyond explicitly discussed items. These findings establish \vogue as both a unique resource for studying multimodal conversational systems and a challenge dataset beyond the current recommendation capabilities of existing top-tier multimodal foundation models such as 
\textsc{GPT-5-mini} and \textsc{Gemini-2.5-Flash}.
\end{abstract}


\maketitle

\section{Introduction}

\begin{figure}[!t]
  \centering
  \includegraphics[width=\linewidth]{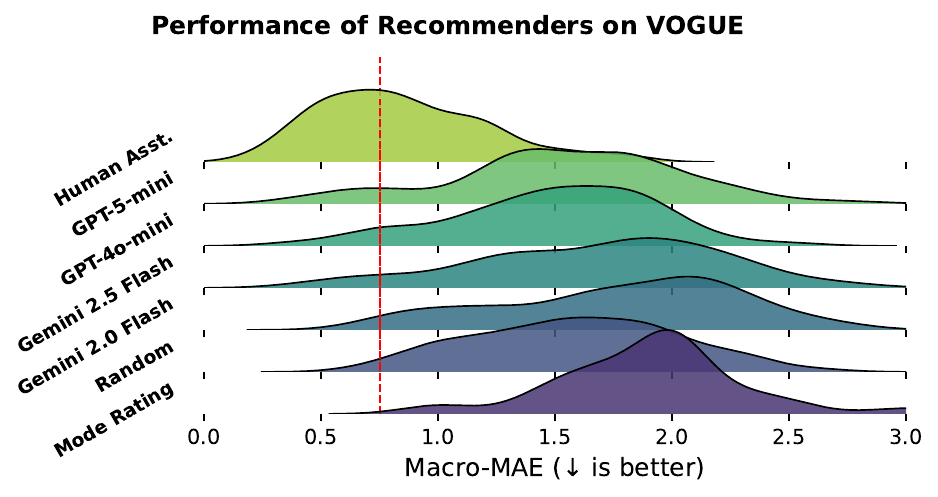}  \caption{Using \vogue, we evaluate a variety of MLLMs on their ability to model user preferences against Human Assistants and two baselines: randomly sampled ratings, and mode (i.e. most common) ratings. Our results show that state-of-the-art MLLMs substantially underperform human agents in preference generalization and understanding.}
  \label{fig:mmae_by_model}
  \Description{Box and Whisker of M-MAE between models, Assistants outperform the models by at least 0.5 on average.}
\end{figure}

\begin{figure}[!b]
  \centering
  \begin{subfigure}[b]{0.45\linewidth}
    \centering
    \includegraphics[width=\linewidth]{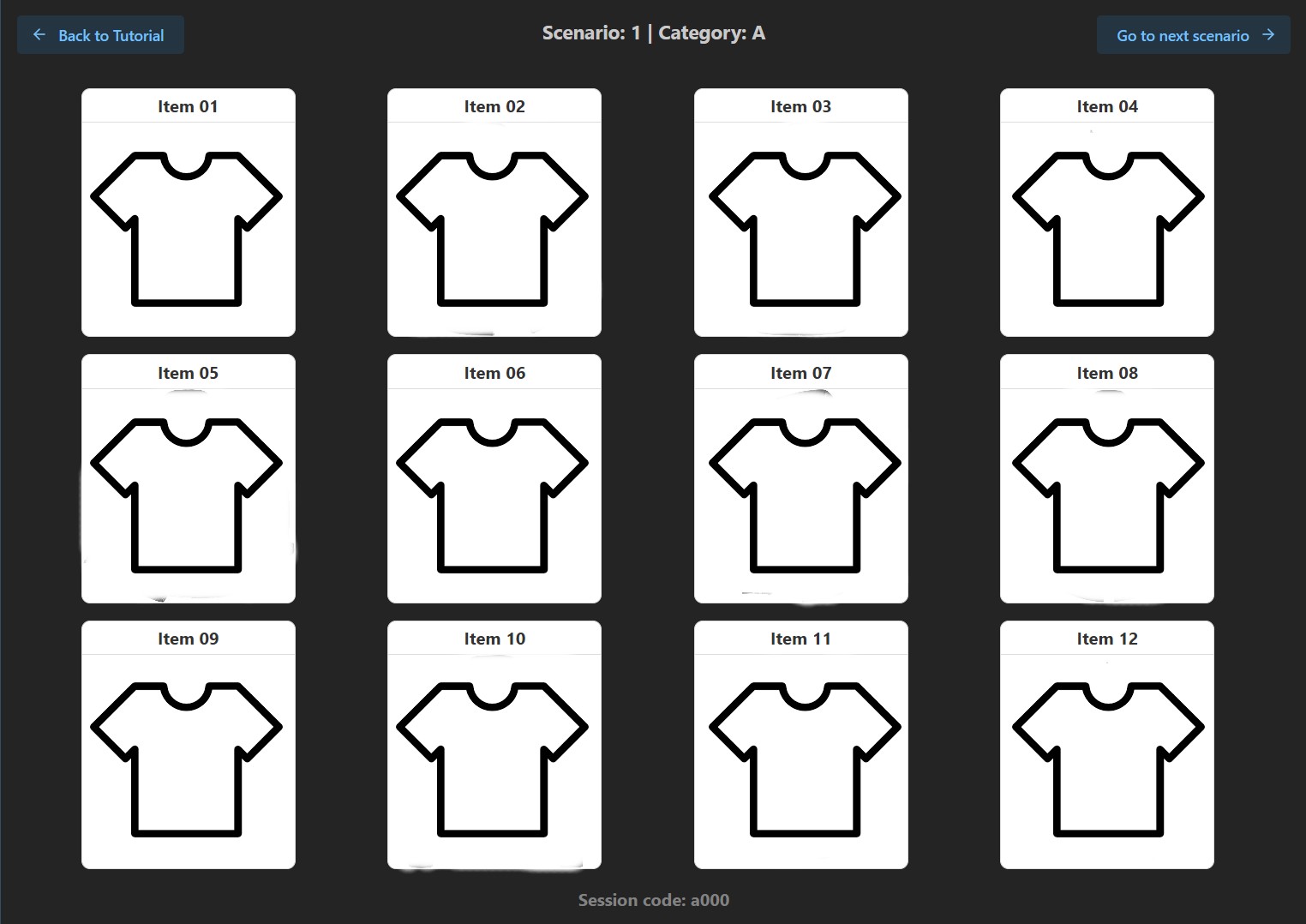}
    \caption{Seeker User Interface}
    \label{fig:seeker_ui}
  \end{subfigure}
  \hfill
  \begin{subfigure}[b]{0.45\linewidth}
    \centering
    \includegraphics[width=\linewidth]{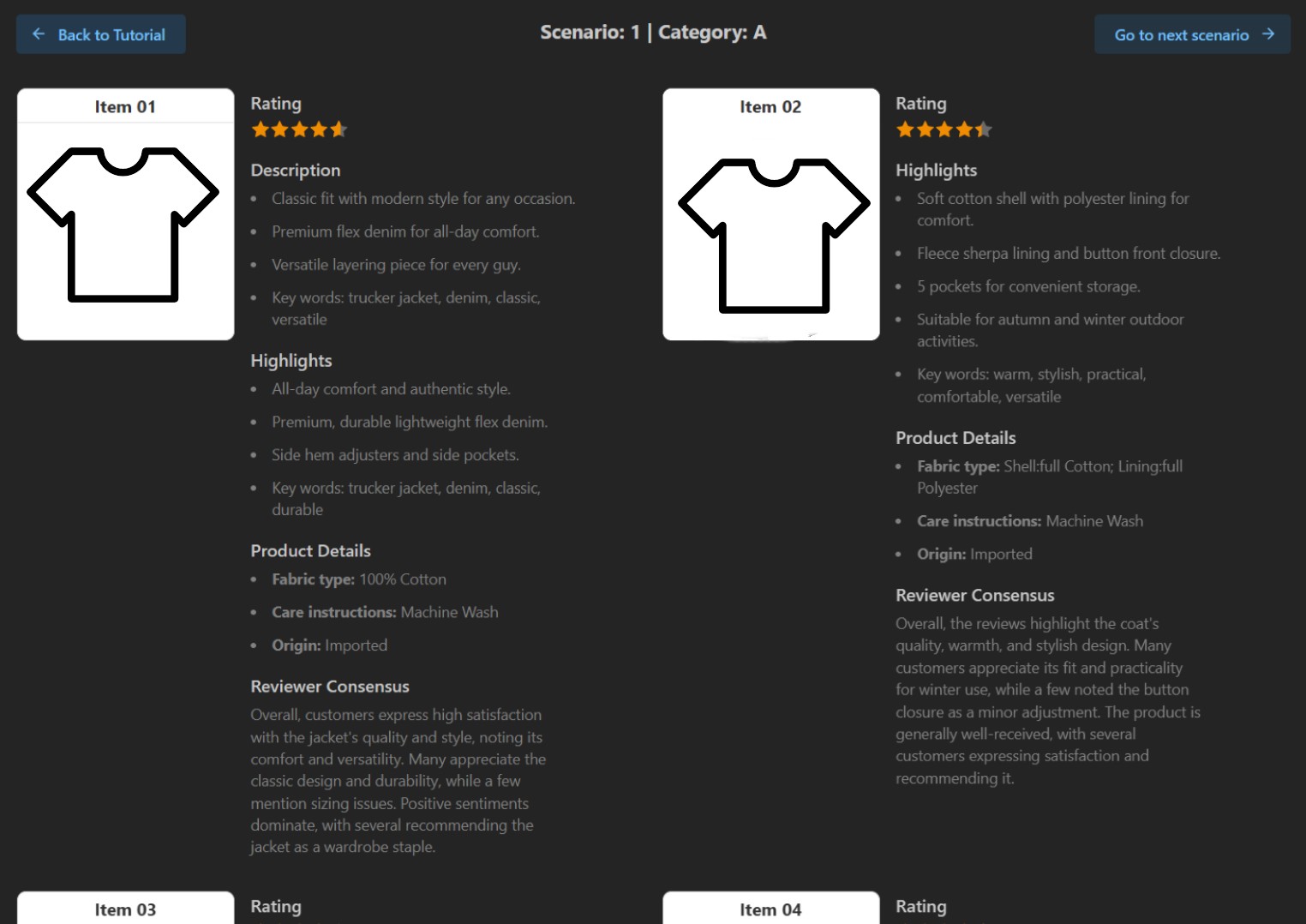}
    \caption{Assistant User Interface}
    \label{fig:asst_ui}
  \end{subfigure}
  \caption{Study User Interfaces used in our experiment. }
  \Description{Images of both UIs used in the study. The Seeker view is a 3x4 grid of product photos, the Assistant view is a 2-column spread of product photos and item descriptions, reviews etc.}
  \label{fig:ui}
  
\end{figure}

Recommender systems have become a central part of the modern Web, shaping how users navigate, consume, and engage with online content across e-commerce, media, and social platforms. As the Web continues to integrate multimodal interaction and human-centered personalization, Conversational Recommender Systems (CRS) have emerged as an expressive paradigm for enabling natural, context-aware user experiences~\citep{chris, Jannach_2021}. However, building effective CRS remains a significant challenge, owing to their inherently subjective nature, requisite multimodality \& human-like conversational ability, and need for generalizable user preference understanding~\citep{GAO2021100}. Among various domains, fashion shopping particularly exemplifies these challenges: preferences are highly subjective and often hinge on visual style and fashion sense, which are difficult to capture through text alone. These issues compound, making a comprehensive evaluation particularly challenging. 

Existing conversational recommendation datasets often fall short in evaluating CRS on such difficulties, 
including
the fashion shopping domain. Most are limited to: a single modality (typically text)~\citep{TG-ReDial, redial,durecdial}; simulated interactions that rely heavily on synthetic rather than organic human dialogues~\citep{sure,MUSE,MMD}; narrow domains such as film recommendation~ \citep{redial,CRM}; or omit personal user profiles~\citep{redial, INSPIRED, MMConv,mgshopdial,simmc2.0, sure, MMD}. Additionally, Large Language Models (LLMs) and Multimodal LLMs (MLLMs) now serve as the backbone of many CRSs~\citep{zhang-2024}, offering strong capabilities for interpreting free-form dialogue and generating natural responses. However, existing datasets and metrics largely emphasize surface-level outcomes, such as the final recommended item, while neglecting the explicit item-level preference ratings and user satisfaction signals~\citep{mgshopdial, MMConv, MUSE}. 
As a result, they fail to expose the limitations of MLLMs compared to human assistants and hinder robust, system-to-system comparison.

In this work we introduce \textbf{\vogue} (\textbf{V}isual-recommendation dial\textbf{O}gue with \textbf{G}rounded \textbf{U}ser \textbf{E}valuations), a dataset that addresses these gaps by featuring (1) multimodal grounding with both visual and verbal inputs, (2) authentic human-human dialogues in realistic fashion shopping settings (Fig.~\ref{fig:ui}), (3) a broad and open-ended fashion domain, and (4) comprehensive user profiles capturing style history, preferences, and ratings of non-candidate items. Each conversation is visually grounded by a catalog of candidate items (including images and metadata) and annotated with full post-conversation item ratings from both the Seeker (user) and Assistant (recommender), and subjective satisfaction metrics captured through post-task Likert surveys. This design enables evaluation along two complementary dimensions: item-level preference alignment and subjective satisfaction metrics. Each conversation is also tagged at the utterance level with a novel intent taxonomy specific to the fashion shopping domain. These measures move beyond accuracy alone, supporting richer and more detailed evaluations across a broad range of conversations.

By releasing \vogue, we aim to provide a practical benchmark (cf. Fig.~\ref{fig:mmae_by_model}) for studying multimodal conversational recommendation. Our dataset establishes an invaluable resource for designing recommendation agents as CRS move into multimodality, providing a foundation to study how dialogue, visual grounding, and user-centered evaluation interact in shaping effective systems.

Our contributions are threefold:

\begin{enumerate}[noitemsep, topsep=0pt]
    \item \textbf{Open Dataset of Real Human Dialogues:} We introduce the first multimodal conversational recommendation dataset in the fashion shopping domain, containing organic human-human dialogues, associated item images and metadata, participant profiles, and detailed rating signals. The dataset provides utterance-level annotations of Assistant and Seeker intents across conversations.
    \item \textbf{Novel Analysis and Insights:} We provide a novel analysis
    for modeling dialogue dynamics, including stage-wise segmentation (Fig.~\ref{fig:vogue_teaser}) of conversations and intent-tagged utterance flows in the multimodal setting.
    \item \textbf{MLLM Benchmark Evaluation:} We benchmark MLLMs against humans on \vogue, revealing persistent challenges in calibration, generalization, and inference, thus positioning it as a challenge dataset for advancing MLLM agents.  
\end{enumerate}

\section{Related Work}

To develop an effective end-to-end learning methodology for multimodal CRS, as well as to establish benchmark datasets for evaluation, studies have contributed meaningful datasets to support recommendation tasks in various domains, as listed in Tab.~\ref{tab:dataset_comparison}. However, existing conversational recommendation datasets are limited and fall short of meeting the needs of our study as follows:
\begin{itemize}
    \item \textbf{Lack of Fashion Datasets:} Most existing conversational recommendation datasets are not about fashion shopping. They usually focus on movies~\citep{redial,TG-ReDial, INSPIRED}, while some cover restaurants, product search/recommendation or travel~\citep{mgshopdial,correccurated, MMConv,wen2024elaborativesubtopicqueryreformulation,wen2025simpleeffectiveelaborativequery}, which are not transferable to our study.
    
    \item \textbf{Lack of Explicit User Feedback:} The rating included along with the existing datasets serves only as an indicator of conversation quality rather than feedback on recommended items during conversations, as well as overall user satisfaction throughout the whole experience    ~\citep{MMConv,correccurated}, which hinders comprehensive evaluation of recommendation quality.
    
    \item  \textbf{Lack of User History and Profiles:} Most existing multimodal conversational datasets either lack personal profiles~\citep{MMD, sure, simmc2.0}, adapt them directly from interaction data~\citep{fashionrec}, or generate them using LLMs~\citep{MUSE}. This reflects a lack of a real user profile associated with the speaker. This limits the opportunity to develop stronger elicitation methods, since long-term preferences are missing~\citep{Jannach_2021}.
    
    \item \textbf{Lack of Real User Dialogue:} For current multimodal conversational datasets targeting fashion shopping, they curated dialogue datasets using data generation methods (e.g., dialogue simulator, LLMs) to produce synthetic dialogue data~\citep{MUSE, fashionrec}. However, the deficiency of realistic conversational behaviors present in the synthetic datasets limits the opportunities to study dialogue patterns.
\end{itemize}

Thus, we believe there is a need for a multimodal conversational dataset supporting visual grounding, consisting of organic human-human dialogues with authentic personal profiles and ratings on recommended items, conversation qualities, and user satisfaction.

\begin{table*}[h]
\centering
\begin{small}
\caption{We present a comparison of existing conversational datasets. Notably, \vogue introduces text and image multimodality, combined with real user profiles, real human dialogue, and comprehensive preference \& user satisfaction feedback.    \textsuperscript{1} Profile is synthetic, created using scraped online data sources. 
    \textsuperscript{2} Profile completed by participant.}
\begin{tabular}{l l l l l c c p{0.8cm}}
\toprule
Name & Domain & Modality & User Profile & Dialogue Source &  Satisfaction Rating & Preference Rating \\
\midrule
DuRecDial 2.0 \citep{durecdial}  & General & Text Only                & Synthetic & Human--Human &\xmark &\xmark  \\ 
ReDial~\citep{redial}      & Film   & Text Only                & None      & Human--Human  & \xmark &\cmark\\
TG-ReDial ~\citep{TG-ReDial}  & Film   & Text Only                & Scraped   & Human--Machine & \xmark &\xmark  \\
INSPIRED ~\citep{INSPIRED}  & Film   & Text Only                & None      & Human--Human & \xmark &\cmark \\
CRM~\citep{CRM}        & Restaurant & Text Only             & Scraped   & Synthetic   &\xmark &\xmark  \\ %
MMConv~\citep{MMConv}     & Travel  & Text + Real Images       & None      & Human--Human & \cmark &\xmark  \\
MG-ShopDial~\citep{mgshopdial} & Product Rec. & Text + Real Images                & None      & Human--Human &\xmark &\xmark\\
CoSRec-Curated~\citep{correccurated} & Product Rec. & Text Only     & Synthetic \textsuperscript{1} & Synthetic &\xmark &\xmark \\
SIMMC 2.0~\citep{simmc2.0}  & Fashion, Furniture & Text + Synthetic Images & None & Human--Human  &\xmark &\xmark\\
SURE~\citep{sure}        & Fashion, Furniture & Text + Synthetic Images & None & Synthetic &\xmark &\xmark\\
MMD~\citep{MMD}        & Fashion & Text + Real Images       & None      & Synthetic    &\xmark &\xmark \\
MUSE~\citep{MUSE}      & Fashion & Text + Real Images       & Synthetic & Synthetic    &\xmark &\xmark  \\
FashionRec~\citep{fashionrec}  & Fashion & Text + Real Images       & Scraped & Synthetic &\xmark &\xmark \\
\midrule \midrule
\textbf{\vogue} & \textbf{Fashion} & \textbf{Text} + \textbf{Real Images}     & \textbf{Self-reported \textsuperscript{2}} & \textbf{Human--Human} &\cmark &\cmark \\
\bottomrule
\end{tabular}
\label{tab:dataset_comparison}
\end{small}
\end{table*}

\section{Dataset Construction}

Based on previously described deficiencies of related work, we aim to curate a novel dataset that satisfies the following desiderata:
\begin{enumerate}
    \item \textbf{Multimodal grounding:} Combine text, images, and metadata to situate dialogues in a realistic setting. 
    \item \textbf{Fashion domain focus:} Address the lack of existing datasets centered on fashion shopping.  
    \item \textbf{Explicit user feedback:} Provide comprehensive and quantitative item-level ratings as well as subjective satisfaction.  
    \item \textbf{User history and profiles:} Include baseline preferences, rationale, and opinions to support personalization.  
    \item \textbf{Organic human--human dialogues:} Collect real conversations rather than synthetic or simulated ones.   
\end{enumerate}

To meet these desiderata, we designed \vogue as six independent conversational recommendation trials between paired participants. We refer to the recommender as the \textbf{Assistant} and the user as the \textbf{Seeker}. Each trial centered on one of 6 unique and orthogonal fashion scenarios, giving the Seeker a temporary decision-making goal. The full scenarios are included in the dataset repository\footnote{The complete \vogue dataset is available at \href{https://github.com/D3Mlab/vogue}{https://github.com/D3Mlab/vogue} \label{dataset_link}}.

In each scenario, participants shared a catalog of 12 candidate items drawn from Amazon.ca, spanning outerwear, layering pieces, and shoes. Each item included images and collated metadata (name, description, review score, review summary). Visual grounding is intrinsic to fashion recommendation, since many critical attributes such as style, silhouette, and overall aesthetic are appearance-dependent. As a result, the task is inherently multimodal and cannot be reduced to a purely textual setting without changing its nature.

To reduce cognitive load, descriptions and reviews were summarized by \textsc{GPT-4o-mini}, while size, fit, color, and price were omitted, and participants were asked not to anchor on them. 

Throughout the study, the catalog comprised 36 items (12 per scenario) plus 40 additional items used for fashion profiles. The Seeker’s task was to articulate needs and select an item, while the Assistant guided preference elicitation and made recommendations. 

\subsection{Protocol}\label{sec:protocol}

Due to space limitations, we present a concise description of the protocol and annotation schema in the paper, while the full experimental protocol, survey instruments, intent definitions, and annotation guidelines are provided in the \textbf{repository}\footref{dataset_link}.  Participant pool and number of trials were set to the largest feasible within our computational and personnel resource budget.

Our data collection protocol\footnote{The protocol received approval from the University of Toronto Research Ethics Board
(\#00049065).} proceeded in three steps:
\begin{enumerate}[noitemsep, topsep=0pt]
    \item \textbf{Onboarding surveys:}\label{sec:onboarding_surveys} All participants completed demographic forms and fashion entrance surveys, including a profile questionnaire and pre-ratings of 40 non-candidate items, which provided baseline preferences and served as user profiles, a feature rarely included in similar datasets. (Desiderata 1, 2, 4). The complete details for all surveys are in the \texttt{surveys} folder of the supplementary material\footref{dataset_link}.
    \item \textbf{Conversational trials:} In each of the six scenarios, the Seeker and Assistant discussed 12 candidate items, with the Assistant accessing full metadata and the Seeker seeing only images and the scenario description. A web-based User Interface (UI) facilitated catalogue viewing, with corresponding Seeker and Assistant views (Fig.~\ref{fig:ui}). Assistants were instructed not to use other external resources and to elicit context through dialogue. Conversations continued until the Seeker indicated satisfaction with at least one item or rejected all options (Desiderata 5). The complete details are in the \texttt{item\_ratings} sub-folder of \vogue\!\footref{dataset_link}.
    \item \textbf{Post-conversation Surveys:} \label{sec:item_rating_surveys} After each scenario, the Seeker provided ground-truth preference ratings of all 12 items on a 1 to 5 scale, while the Assistant attempted to predict these ratings. Participants were instructed to rate based on how likely they would be to purchase the item given the conversation and scenario, with 5 being “would purchase” and 1 being “would never purchase”. Neither party saw the other's ratings. Both parties also completed a short feedback survey on the perceived quality of that conversation. Seekers additionally completed a 1-5 Likert scale survey of their satisfaction with the scenario and conversation. Survey questions are detailed in Tab.~\ref{tab:likert_list} (Desiderata 3).
\end{enumerate}
\begin{table}[!t]
\centering
\caption{Seeker satisfaction questions (1--5 Likert scale).}
\label{tab:likert_list}
\begin{tabular}{p{0.95\linewidth}}
\toprule
\begin{itemize}[leftmargin=*, nosep]
    \item The conversation allowed me to make a good selection on what to buy.
    \item I made a well-justified final selection.
    \item I would actually purchase the item selected.
    \item During the conversation, I felt in control of the process.
    \item I found the Assistant's recommendations helpful.
\end{itemize}
\\
\bottomrule
\end{tabular}
\end{table}

\subsection{Participants}
We recruited 20 participants through informal outreach at the University of Toronto 
and among recent graduates in the surrounding area. Young adults were recruited in order to reflect a core demographic of fashion shopping who are potentially amenable to the use of conversational assistants.  
Participants were randomly assigned fixed roles as Seeker or Assistant for the entire session. To prepare Assistants for their role, we conducted a short coaching session to familiarize them with the catalog of candidate items. Seeker participants received only the task scenarios beforehand to replicate real-world conditions, necessitating that Seekers articulate their needs without access to full product knowledge.

\subsection{Dialogue Intent}\label{sec:dialogue_intent}
To better understand dialogue patterns with visual support in fashion shopping recommendations, we extended the holistic user and intent definitions from Lyu et al.~\citep{zoey} and also referred to the work of Cai and Chen on utterance-level labeling~\citep{caichen}. Since earlier studies focused on domains outside fashion shopping under conversational recommendation \emph{without} visual grounding in conversations, we created new taxonomies on both sides to better categorize dialogue intents and allow for more thorough analysis. The taxonomies used in this study, designed for dialogue actions in a fashion shopping setting with shared visual catalogue, are listed in Tab.~\ref{tab:intent_taxonomy}.

We conducted a two-phase labeling process to annotate dialogue utterances for subsequent analysis. In the first phase, we leverage \textsc{Gemini-2.5-Flash} to streamline large-scale tagging. In the second phase, all tagged conversations were manually reviewed by human annotators to correct mislabeled utterances and to refine and expand the taxonomy based on the collected dialogues. Pairwise inter-annotator reliability was assessed using Cohen’s Kappa, yielding a strong agreement of $\kappa = 0.81$, which indicates high consistency in intent labeling. From these tags, we derive the intents used in Sec.~\ref{sec:stage}. For clarity, we primarily discuss overarching intent categories (e.g., \textit{Provide} or \textit{Explain}) rather than specific tags. 

\begin{table*}[t]
\centering
\begin{footnotesize}
\caption{Intent taxonomy, building on \citet{zoey, caichen}. 
Tags marked with $^\dagger$ are newly introduced in \vogue.}
\label{tab:intent_taxonomy}
\begin{tabular}{@{}>{\raggedright\arraybackslash}p{0.22\textwidth} >{\raggedright\arraybackslash}p{0.42\textwidth} >{\raggedright\arraybackslash}p{0.30\textwidth}@{}}
\toprule
\textbf{Category (Code)} & \textbf{Description} & \textbf{Example} \\
\midrule

\multicolumn{3}{@{}l@{}}{\normalsize\bfseries Seeker} \\
\addlinespace[0.2em]
\multicolumn{3}{@{}l@{}}{\textbf{Ask for Recommendation}} \\
\hspace{0.5em}Initial Query (IQ) & ...asks for a recommendation in the first query & ``Hey, I want to purchase one coat...'' \\
\hspace{0.5em}Continue (CON) & ...asks for another recommendation & ``Would you give me a Plan B instead?'' \\

\multicolumn{3}{@{}l@{}}{\textbf{Provide}} \\
\hspace{0.5em}Provide Context (PCT) & ...provides background or situational information & ``I want to choose outerwear for a fall visit to a farm...'' \\
\hspace{0.5em}Provide Explicit Preferences (PEP)$^\dagger$ & ...states explicit constraints or desired attributes & ``It has to be waterproof.'' \\
\hspace{0.5em}Provide Implicit Preference (PIP)$^\dagger$ & ...reveals preferences indirectly through reactions or comparisons & ``I don’t really like Item11 compared to Item10.'' \\
\hspace{0.5em}Refine Preference (RP) & ...refines or relaxes previously stated preferences & ``I don’t need a hood necessarily...'' \\
\textbf{Answer (ANS)} & ...answers a question from the Assistant & ``The weather is cool with light rain.'' \\
\textbf{Acknowledgement (ACK)} & ...shows understanding or confirmation & ``Yeah.'' \\

\multicolumn{3}{@{}l@{}}{\textbf{Recommendation Rating}} \\
\hspace{0.5em}Interest (INT)$^\dagger$ & ...reacts positively without committing to a final choice & ``I do like Item08 and Item05.'' \\
\hspace{0.5em}Accept (ACT) & ...accepts the recommended item & ``I think I’ll go with Item02.'' \\
\hspace{0.5em}Reject (RJT) & ...rejects the recommendation & ``Item03 is too ugly for me.'' \\
\hspace{0.5em}Neutral Response (NR) & ...does not indicate acceptance or rejection & ``I see.'' \\

\multicolumn{3}{@{}l@{}}{\textbf{Inquire}} \\
\hspace{0.5em}Inquire -- Factual (IF)$^\dagger$ & ...asks for factual information about an item & ``Does it have a pocket?'' \\
\hspace{0.5em}Inquire -- Ask Opinion (IA)$^\dagger$ & ...asks for the Assistant’s subjective opinion & ``Which one do you think is more comfortable?'' \\

\multicolumn{3}{@{}l@{}}{\textbf{Critiquing}} \\
\hspace{0.5em}Critique -- Feature (CF) & ...critiques a specific attribute of an item & ``Item05 is a bit too long.'' \\
\hspace{0.5em}Critique -- Compare (CC) & ...requests comparison between items & ``Between Item23 and Item16, which is better?'' \\

\textbf{Explain (EXP)$^\dagger$} & ...explains reasons behind acceptance, rejection, or preference & ``I like Item05 because it goes well with my outfit.'' \\
\textbf{Ask Clarification (AC)} & ...asks clarification of something previously said & ``Did you say Item13 as well?'' \\
\textbf{Others (OTH)} & Greetings, gratitude, or chit-chat & ``Hello.'' \\

\midrule

\multicolumn{3}{@{}l@{}}{\normalsize\bfseries Assistant} \\
\addlinespace[0.2em]

\multicolumn{3}{@{}l@{}}{\textbf{Request}} \\
\hspace{0.5em}Request Task Initiation (RTI)$^\dagger$  & ...initiates dialogue with an open-ended prompt & ``How can I help you today?'' \\
\hspace{0.5em}Request Preferences (RP)$^\dagger$ & ...elicits the Seeker’s preferences or constraints & ``Are you looking for something longer?'' \\
\hspace{0.5em}Request Context (RC)$^\dagger$ & ...elicits situational context & ``Where is this farm going to be?'' \\
\hspace{0.5em}Clarify Question (CQ) & ...asks clarification of a prior requirement & ``Do you mean a more formal setting?'' \\
\hspace{0.5em}Ask Opinion (A) & ...asks the Seeker to choose or judge among options & ``Which one do you prefer?'' \\
\hspace{0.5em}Ensure Fulfillment (EF) & ...checks whether the task is complete & ``Any other items you want to see?'' \\

\textbf{Inform Progress (IP)} & ...indicates processing or navigation state & ``Give me one second, I’m checking Item29...'' \\
\textbf{Acknowledgement (ACK)} & ...acknowledges a Seeker utterance & ``Got it.'' \\
\textbf{Answer (ANS)} & ...answers a Seeker’s question & ``Waterproof is stronger than water-resistant.'' \\

\multicolumn{3}{@{}l@{}}{\textbf{Recommend}} \\
\hspace{0.5em}Recommend -- Show (RS) & ...presents candidate items & ``Take a look at Item31, Item32, and Item28.'' \\
\hspace{0.5em}Recommend -- Combine (REC)$^\dagger$ & ...recommends compatible combinations or layered outfits & ``You can wear Item22 under Item17.'' \\

\multicolumn{3}{@{}l@{}}{\textbf{Explain}} \\
\hspace{0.5em}Explain -- Preference (EP) & ...justifies recommendations using preferences & ``Item03 is best because it’s waterproof.'' \\
\hspace{0.5em}Explain -- Additional Info (EAI) & ...adds new factual or descriptive details & ``It’s machine washable.'' \\

\multicolumn{3}{@{}l@{}}{\textbf{Personal Opinion}} \\
\hspace{0.5em}Comparison (PCM) & ...compares two or more items & ``Only Item11 is machine washable.'' \\
\hspace{0.5em}Persuasion (PER) & ...adds positive framing or encouragement & ``Both materials are pretty good.'' \\
\hspace{0.5em}Prior Experience (PEX) & ...references personal experience & ``I have one myself, it’s my go-to boot.'' \\
\hspace{0.5em}Context (PCN) & ...gives opinion grounded in context & ``This matters more in deep snow.'' \\

\textbf{Others (OTH)} & Greetings, gratitude, or chit-chat & ``Hi.'' \\
\bottomrule
\end{tabular}
\end{footnotesize}
\end{table*}

\begin{table}[t]
\begin{footnotesize}
    
\centering
\caption{Vogue dataset statistics.}
\label{tab:dataset_stats}
\begin{tabular}{l r l r}
\toprule
Statistic & Count & Statistic & Count \\
\midrule
Conversations & 60 & Participants & 20 \\
Total turns & 899 & Total utterances & 2,100 \\
Total tokens & 22,717 & Total item ratings & 1,440 \\
Avg.\ turns / conversation & 14.8 & Avg.\ tokens / turn & 36.1 \\
\bottomrule
\end{tabular}
\end{footnotesize}
\end{table}

\subsection{Dataset Contents}

\vogue contains 60 two-party dialogue transcripts, transcribed and time-stamped at the turn level. \textbf{All 2{,}100 utterances are labeled} with a comprehensive intent taxonomy for both roles. Transcripts are additionally annotated with final item choices and catalogue items explicitly mentioned, and each scenario is paired with its corresponding 12-item visual catalogue released with item metadata. 

Beyond transcripts, \vogue includes dense preference supervision: post-scenario item-rating surveys over all items from both Seeker (ground truth) and Assistant (predicted), yielding \textbf{1{,}440 total item-level ratings}, plus post-task satisfaction  surveys. We additionally collected fashion-profile surveys for all participants to contextualize preference priors. We position \vogue as a \textbf{high-fidelity evaluation and analysis benchmark} for multimodal CRS, and as a challenge dataset exposing shortcomings in visual grounding, preference inference, and preference generalization.

All collected data (cf. Sec.~\ref{sec:protocol}) are available in \vogue's repository\!\footref{dataset_link},  
including user profiles (\texttt{fashion\_profile} folder), item rating surveys from both Seeker and Assistant (\texttt{surveys} folder), and conversational transcripts (\texttt{conversations} sub-folder of \texttt{data} folder). Major dataset statistics are listed in Tab.~\ref{tab:dataset_stats}.

\section{Data Analysis}

In this section, we examine CRS dialogues from both a qualitative and quantitative perspective, considering their overall structure, the alignment between Seeker and Assistant preferences, and the performance of modern MLLMs relative to the human Assistants. 

Specifically, we focus our analysis with \vogue on three complementary aspects: \textbf{visual grounding}, which examines how shared images shape dialogue dynamics in inherently multimodal fashion scenarios; \textbf{preference alignment}, which tests whether Assistants' alignment with both direction and intensity of Seeker preferences improves Seeker satisfaction; and \textbf{MLLM as recommender}, which evaluates whether current models can move beyond surface-level dialogue cues toward robust and generalizable preference modeling. 

To this end, we propose three Research Questions (RQs):

\begin{itemize}

    \item[\textbf{RQ1}:]\textbf{Visual grounding \& dynamics.} 
    How does visual grounding alter the dynamics and stages of dialogue, compared to text-only settings? In an inherently multimodal domain like fashion, what structures and processes emerge, and how should CRS adapt to these?

    \item[\textbf{RQ2}:]\textbf{Impact of Human Assistant alignments on Seeker experience.} 
    To what extent do the item-level preference ratings predicted by Assistants after a conversation match the ground-truth ratings from Seekers? And does this alignment on preference actually correlate with the subjective satisfaction of Seekers?
    
    \item[\textbf{RQ3}:]\textbf{MLLM versus Human Assistant Alignment.} 
    How well do MLLMs perform when they are asked to recover a Seeker's ratings given the conversation? What strengths and weaknesses do they show compared to human Assistants? 

\end{itemize}

By framing our analysis around these questions, we highlight how \vogue both enables deeper evaluation of existing systems and suggests design principles needed for user-centric multimodal CRS.

\subsection{RQ1: Visual grounding \& dynamics.} \label{sec:stage}

To address \textbf{RQ1}, we examine how dialogue progresses in an inherently visual recommendation setting, and what structural and behavioral changes result from this grounding. Since a direct non-visual ablation is not possible in this domain, we examine the distribution of conversational intent, item mentions, and intent tag transitions across the course of the interaction and compare with related datasets. For the sake of clarity, our analysis will omit \textit{Acknowledgment} and \textit{Others} tags, since they do not meaningfully advance the conversation.

\begin{figure}
    \centering
\includegraphics[width=0.9\linewidth]{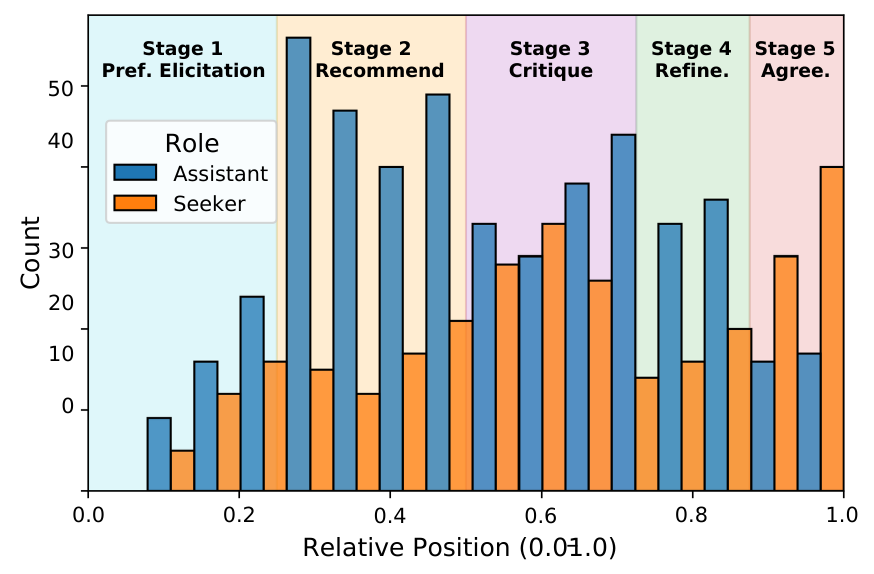}
    \caption{Seeker \& Assistant Item Mention Distribution by Conversation Proportion. We observe a bimodal distribution in the Assistant, corresponding to \textit{First Recommendation} and \textit{Refinement} stages respectively.}
    \label{fig:sa_utt_dist}
    \Description{Seeker \& Assistant Item Mention Distribution by Conversation Proportion. Stages are also labeled. Assistants show a strongly bi-modal twin peak shape which correspond with Stage 2 and 3, while Seekers showed peaks in Stage 3 and 5}
\end{figure}

As shown in Fig.~\ref{fig:sa_utt_dist}, Assistants consistently introduce new items in two distinct waves: the first following preference and context elicitation, and the second after refinement based on earlier feedback. We also observe that Assistants frequently group recommendations, presenting up to four items simultaneously for discussion. This contrasts with conversational flows in text-only datasets, where item presentation tends to be more linear and sequential. Chiefly, we observe that the conversations flow naturally through a common implicit procedure shaped by visual grounding.

\begin{figure*}
  \centering
  \includegraphics[width=0.90\linewidth]{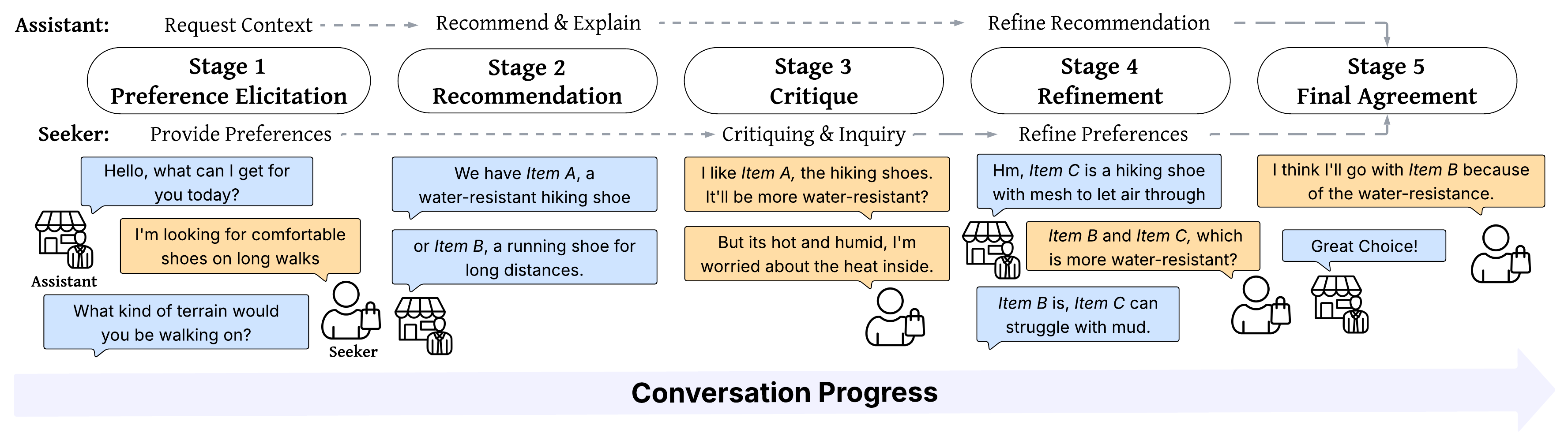}
  \caption{Our proposed stage splits, illustrating Seeker-Assistant initiative and major intent tags. Examples of dialogue are included, along with each role's motivation at each stage.}
  \label{fig:vogue_teaser}
  \Description{A 5-stage flowchart/summary of the stage analysis.}
\end{figure*}

The distributions in Fig.~\ref{fig:sa_utt_dist} further suggest a clear stage-like structure: activity is clustered into distinct peaks rather than being evenly spread across the dialogue, with a characteristic ebb and flow as initiative flips between participants. We therefore segment the conversations into five stages, each defined by the dominant (i.e., most prevalent) intents and role dynamics at that point. In Fig.~\ref{fig:vogue_teaser}, we provide an illustrative example of each stage based on conversations in \vogue: 

\begin{enumerate}\leftskip12pt
    \item[\textbf{Stage 1:}] \textbf{Preference Elicitation.} Early exchanges are dominated by preference elicitation (Assistant \textit{Request}; Seeker \textit{Provide}, \textit{Ask for Recommendation}, and \textit{Answer}).
    \item[\textbf{Stage 2:}] \textbf{Recommendation.} This is followed by the first recommendation wave (Assistant \textit{Recommend}), with \textit{Explain} rising soon after.  Interestingly, we observed that participants often discussed the items in groups according to abstract unifying properties of those item groups.
    \item[\textbf{Stage 3:}] \textbf{Critique.} The conversation transitions to a bridge characterized by \textit{Critique}, \textit{Inquire}, and evaluative ratings, while Assistant \textit{Explain} and \textit{Personal Opinion} remain high.
    \item[\textbf{Stage 4:}] \textbf{Refinement.} Stage 4 is strongly characterized by Assistants introducing a new, reduced set of items (usually one or two) to refine the recommendation based on Seeker critiques. Concurrently, Seekers refine their preferences based on available item attributes, continuing to utilize property grouping as observed in \textit{Stage 2}.
    \item[\textbf{Stage 5:}] \textbf{Final Agreement.} Finally, Stage 5 is marked by Seeker critique and \textit{Recommendation Rating} peaking, particularly \textit{Accept} tags, signaling choice and resolution.
\end{enumerate}

\paragraph{\textbf{Significance to CRS}}
Our analysis indicates that visual grounding and multimodality reshape conversational dynamics compared to text-only CRS settings. Dialogues in \vogue\ tend to follow a staged structure, supported by shared images that keep exchanges relatively compact. Conversations average 14.8 turns and 36 tokens per turn, yet still converge on final agreements with strong preference alignment (see Sec.~\ref{sec:alignment}).

Anecdotal inspection of the transcripts indicates frequent usage of positional identifying stylistic references in the shared catalogue that reduces the need for verbose description or minute back-and-forth; contrasting with similar studies lacking a shared visual catalogue~\citep{zoey}, and allows participants to anchor their discussion and reasoning in concrete visual referents. 

Taken together, these observations point to design implications for CRS: multimodal grounding can enable parallel exploration of grouped items, comparative reasoning, and collaborative refinement, interaction patterns that are far less accessible in text-only dialogue flows. These mechanisms could be directly incorporated into CRS design, providing a way to make interactions more efficient and user-friendly. Exploring how such patterns translate into deployed systems offers a promising direction for future work.

\subsection{RQ2: Human Assistant Alignment and Impact on Seeker experience.}

We now evaluate the quality of human Assistant interaction in \vogue using two complementary metrics: (1) quantitative alignment between human Assistant predictions and Seeker ground-truth ratings, and (2) post-task satisfaction surveys from Seekers. 

Following~\citet{Jannach_2021} and~\citet{mcnee}, we choose alignment as recovery of both the structure and intensity of preferences over accuracy, reporting it via Pearson Correlation (PC) and Mean Absolute Error (MAE) between item-level ratings. Satisfaction is included as an orthogonal but essential user-centered measure and reported on a 1-5 Likert scale. 

\paragraph{\textbf{Alignment between Assistants and Seekers}}
\label{sec:alignment}

\begin{figure}[t]
  \centering
  \includegraphics[width=1\linewidth]{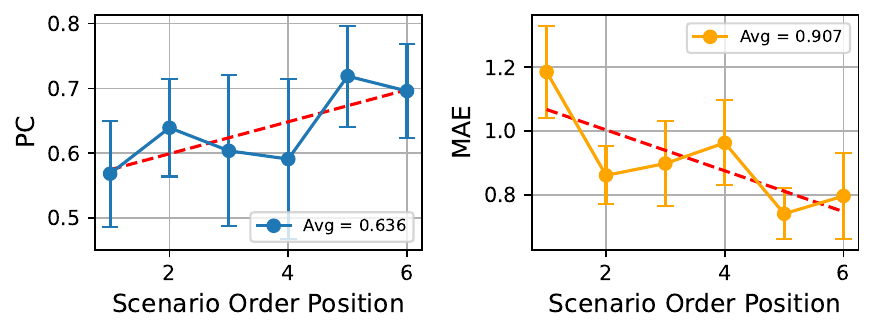}
  \caption{Alignment metrics as a function of scenario order position. Scenario order denotes the order in which individual trials were conducted for each Assistant--Seeker pair. }
  \label{fig:order_metrics}
  \Description{Two side-by-side plots showing alignment metrics by scenario order position. The left plot shows Pearson correlation by scenario order position, and the right plot shows mean absolute error by scenario order position. Scenario order denotes the order in which individual trials were conducted for each Assistant--Seeker pair.}
\end{figure}

We observed that Seekers strongly favored low scores (44\% 1s, only 8.6\% 5s), resulting in a highly imbalanced distribution. Human Assistants nonetheless achieved good preference alignment, with a mean PC of 0.64 (median 0.61, SD 0.21) as shown in Fig.~\ref{fig:order_metrics}, indicating reliable recovery of preference directions rather than isolated success cases. MAE varied more across scenarios, from 0.25 to ~2.0, averaging 0.90, underscoring persistent challenges in modeling preference strength.

Since each Seeker engaged with the same Assistant in six scenarios, Fig.~\ref{fig:order_metrics} also analyzes the performance across repeated interactions. PC improved and MAE declined over successive scenarios, suggesting that Assistants became more effective with practice, despite neither party ever seeing the other's item ratings, likely due to increased dyad-specific calibration over time. While improvement trends were modest, the result underscores the value of long-horizon modeling in CRS and indicates potential for adaptive learning without explicit item-level feedback. 

\paragraph{\textbf{Correlation between Alignment and Satisfaction.}}

Unlike prior CRS resources, our design captures not only objective alignment through item ratings but also subjective satisfaction from Seekers, enabling a more holistic evaluation of recommendation quality. In Fig.~\ref{fig:al_likert_cor}, we observe that Seeker satisfaction (y-axis) was consistently high. Likert surveys captured perceived decision quality, justification, intent to purchase, and Assistant helpfulness. Most scores exceeded 4.5 (out of 5), with helpfulness especially positive.

We further analyze the correlation between the Assistant’s alignment and the Seeker’s satisfaction (cf. Fig.~\ref{fig:al_likert_cor}). Encouragingly, higher satisfaction generally coincides with stronger alignment (measured by PC and MAE), indicating that better alignment contributes to better user experiences. This strong correlation also supports the use of alignment metrics as a proxy for preference inference quality when user satisfaction is unknown, such as in the post-conversation MLLM evaluation in the following sections. 

Nonetheless, satisfaction varied across perceived decision impact and purchase intent, underscoring that it is not monolithic. This suggests that measuring only alignment, as is common in existing datasets, is insufficient. By contrast, the comprehensive design of \vogue highlights the opportunity for CRS to maximize impact by coupling robust preference modeling with user-centered design. In accordance with prior research~\citep{mcnee}, a system that feels supportive and transparent, while also aligning with user preferences, is more likely to foster sustained user satisfaction and trust.

\subsection{RQ3: MLLM versus Human Assistant Alignment.}\label{sec:rq3}
The recent surge of MLLMs in conversational recommendation~\citep{zhang-2024} 
leads to questions about their ability to operate on human-human dialogue in recommender and preference modelling roles. Having established the impact that alignment can have on user experience, we now examine how well MLLMs perform in this domain. To answer \textbf{RQ3}, we provide an evaluation of several prevalent MLLMs on the proposed dataset.  We provide the prompt template as follows:

\begin{figure}
    \centering
    \includegraphics[width=0.9\columnwidth]{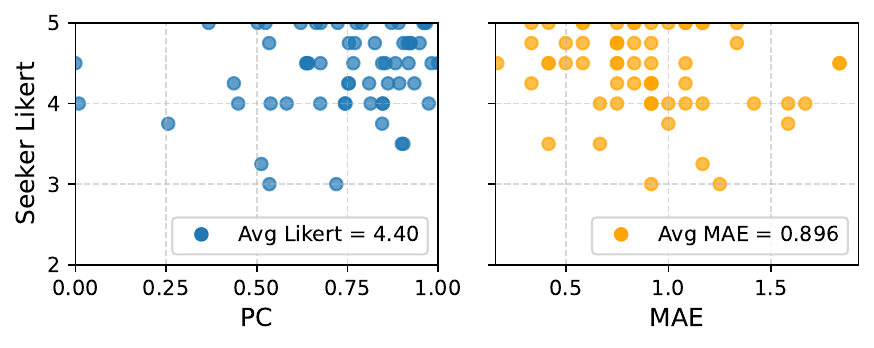}
    \caption{Relationship between Seeker satisfaction and alignment metrics. PC (left) shows strong clustering with higher satisfaction, while MAE (right) reveals broader variability.}
    \label{fig:al_likert_cor}
    \Description{Two scatter plots in a single image: Likert vs. PC on the left, and Likert vs. MAE on the right. Both show satisfaction correlates with alignment.}
\end{figure}

\vspace{0.5em}

{\small\ttfamily
\noindent
You are an expert sales associate and fashion expert, with deep experience in apparel,
styling, sales, and customer relations, reviewing a recorded conversation between an
assistant and a seeker. The assistant provides shopping advice and help to the seeker,
who is attempting to make a clothes purchase. You must rate each of the 12 items on a
1-5 scale (inclusive).

\medskip
\noindent
Criteria: Rate the items based on how likely you think the Seeker is to purchase them,
given the conversation and scenario that just concluded.

\medskip
\noindent
<format guidelines>\\
---\\
<catalogue, metadata and conversation history here>
}

The goal of this experiment is to evaluate how well MLLMs can recover Seeker preferences from the available post-conversation evidence, and not end-to-end conversational policy. We do not compare specific fine-tuned or purpose-built MLLM-based CRS for fashion recommendation, as it is beyond the scope of this work. Accordingly, all models are provided with the same prompt as above, the full item catalogue \& metadata, and the complete tagged transcript at the conversation level. We then instruct them to complete the same item rating survey task as the Assistants. We also include two baselines: (i) \textit{Baseline: Random}, which randomly samples ratings from the Seeker's empirical ground truth distribution, and (ii) \textit{Baseline: Mode Rating}, which returns the mode (i.e. most common) rating of 1. Our analysis first examines rating distributions, followed by item-level alignment and preference generalization.

\paragraph{\textbf{Performance Overview}}

\begin{table*}[t]
\centering
\begin{small}
\caption{Alignment metrics and accuracy for human Assistants and MLLMs. 
Best results from models and Assistants are \textbf{bolded}, second-best are \underline{underlined}. 
Arrows indicate whether higher or lower is better. \textsuperscript{*}PC undefined (uniform rating, zero variance)}
\label{tab:llm_stats}
\begin{tabular}{l c c c c c c c c c}
\toprule
\textbf{Model} & \textbf{Accuracy ↑} & \textbf{PC ↑} & \textbf{MAE ↓} & \textbf{M-MAE ↓} & 
\footnotesize{\textbf{MAE[GT=1] ↓}} & \footnotesize{\textbf{MAE[GT=2] ↓}} & 
\footnotesize{\textbf{MAE[GT=3] ↓}} & \footnotesize\textbf{{MAE[GT=4] ↓}} & 
\footnotesize{\textbf{MAE[GT=5] ↓}} \\
\midrule
Human Assistants & 0.397 & \textbf{0.636} & \textbf{0.896} & \textbf{0.849} & \textbf{0.746} & \underline{0.891} & \underline{1.171} & \underline{1.284} & \textbf{0.597} \\
\textsc{GPT-5-mini}       & 0.344 & \underline{0.083} & 1.296 & 1.543 & 1.041 & \textbf{0.845} & 1.209 & 2.222 & 2.516 \\
\textsc{GPT-4o-mini}      & 0.156 & 0.000 & 1.619 & \underline{1.491} & 2.160 & 1.186 & \textbf{0.953} & \textbf{1.160} & \underline{1.726} \\
\textsc{Gemini-2.5-Flash} & \textbf{0.406} & 0.057 & \underline{1.356} & 1.681 & \underline{0.853} & 0.930 & 1.628 & 2.444 & 2.839 \\
\textsc{Gemini-2.0-Flash} & \underline{0.402} & 0.041 & \underline{1.356} & 1.772 & 0.693 & 1.031 & 1.736 & 2.568 & 3.065 \\
\midrule
Baseline: Random  & 0.240 & -0.005 & 1.713 & 1.730 & 1.997 & 1.403 & 1.147 & 1.469 & 2.387 \\
Baseline: Mode Rating  & 0.443 & NaN\textsuperscript{*} & 1.219 & 1.932 & 0.000 & 1.000 & 2.000 & 3.000 & 4.000 \\
\bottomrule
\end{tabular}
\end{small}
\end{table*}

As before, we report PC and MAE; however, due to class imbalance (which most strongly affects the MLLMs) we also report per class ($\text{MAE}[\text{GT}=k]$ for Seeker ground truth (GT) rating $k$) and Macro MAE (M-MAE). 
We report accuracy for all models, though we discuss later that this statistic can be misleading due to class imbalance. Due to the retroactive nature of these evaluations, no model is actively participating in the recommendation process so satisfaction metrics are not relevant here. 
The results are presented in Tab.~\ref{tab:llm_stats}. We note that the Assistants score best or 2nd best in nearly every measure, even accounting for the learning effects seen in Sec. \ref{sec:alignment}. \emph{Critically, no MLLM alternative tested was competitive with real human recommenders.}

\paragraph{\textbf{Calibration and Distribution}} \label{sec:calibration}

\begin{figure}
  \centering
  \includegraphics[width=0.8\linewidth]{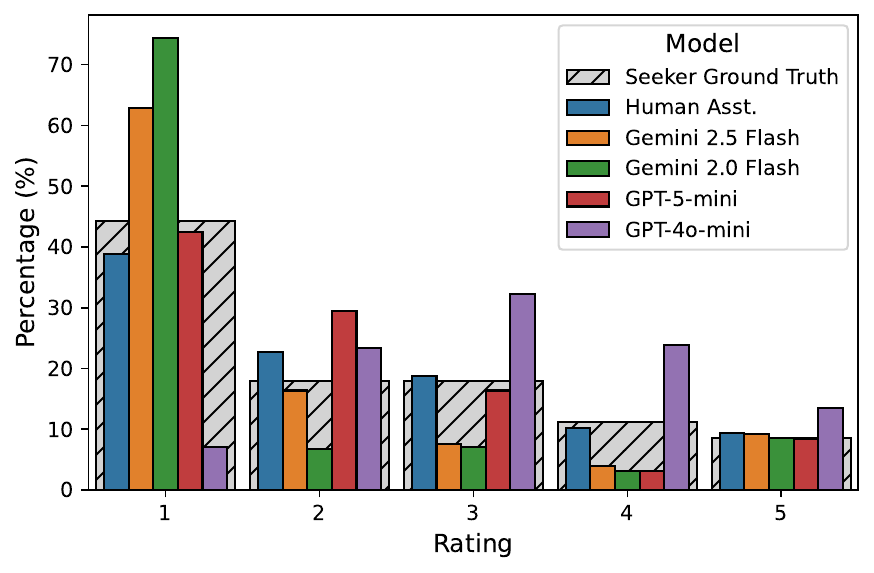}
  \caption{Distribution of Predicted Ratings vs. \ Seeker Ground Truth. Ground truth rating distribution is presented in grey.}
  \label{fig:rating_dist}
  \Description{Ratings distributions for all models. GPT-5-Mini matched best.}
\end{figure}

Examining a distribution of ratings, we note that both families of models experience systemic failure in replicating Seeker ratings; \textsc{Gemini} models heavily over-estimate the proportion of 1 ratings, while heavily collapsing on 3s and 4s; \textsc{GPT-4o-mini} adheres to a strongly normal distribution, peaking at 3, while \textsc{GPT-5-mini} produces a nearly linear decline across ratings (Fig.~\ref{fig:rating_dist}). All models maintain a relatively fair proportion of 5 ratings. This deficit is directly reflected in the PC performance of each model, where they perform very poorly, especially in comparison to human Assistants. Poor performance at capturing natural human rating distributions directly undermines the ability of these models to effectively predict preferences. 

\paragraph{\textbf{Weak Preference Generalization}}

Next we specifically examine how well different recommenders are able to infer and generalize preference knowledge within the conversation. Since it has been shown that most state-of-the-art MLLMs are able to parse explicit dialogue signals well~\citep{pandia}, we foreground the ability to extend and generalize explicit preferences onto the full catalogue. Since many of these items \emph{may not have been explicitly discussed}, the ability to 
generalize from the discussion to accurately predict any item rating
in the full catalogue is paramount for effective recommendation. 

From the per-conversation M-MAE for the MLLM models and two baselines, we note a distinct and consistent gap from the Assistants' performance (cf. Tab.~\ref{tab:llm_stats} and Fig.~\ref{fig:mmae_by_model}). Both Gemini models perform worse than the Random baseline. This underscores that extending preference inference beyond explicitly discussed items  
remains a challenging open problem, even for state-of-the-art MLLMs. 

\paragraph{\textbf{Implications for MLLMs in CRS}}
Viewed from the naive lens of accuracy, 
many MLLMs \emph{appear} to perform comparably to human Assistants. However, this apparent parity is largely an artifact of class imbalance and the relative ease of recovering extreme ratings from conversational cues. When results are disaggregated by rating (i.e., $\text{MAE}[\text{GT}=k]$ in Tab.~\ref{tab:llm_stats}), none clearly surpasses the Assistants. 
In fact, a trivial and uninformed strategy of predicting all 1s (Baseline: Mode Rating) beats all MLLMs in raw accuracy. Accordingly, \emph{accuracy is not a meaningful primary metric} for preference alignment in \vogue; we recommend PC as the headline measure.

This dissonance stems in part from systematic distortions in rating distributions, which prevent MLLM-based recommenders from faithfully capturing Seeker ambivalence or partial fit. In turn, this weakens their ability to generalize preference reasoning beyond explicit conversational cues and actively discussed items. We do not observe substantial evidence that any of these models form a coherent preference model of the Seeker.

In direct response to \textbf{RQ3}, these results demonstrate that extending preference inference beyond explicit conversational cues remains an open challenge for state-of-the-art MLLMs. The dataset introduced here directly foregrounds this gap by enabling evaluation across fully featured candidate catalogues and comprehensive and detailed item-level feedback, providing a benchmark that moves beyond surface-level metrics like accuracy and supports rigorous comparison of MLLM-based CRS systems in terms of their capacity for nuanced and generalizable preference modeling.

\section{Conclusion}
We introduced \vogue, a multimodal conversational recommendation dataset that addresses current gaps by combining (i) \textbf{multimodality}, with fashion catalogs containing both product images and metadata to ground dialogue in visual and textual context; (ii) \textbf{human--human conversations}, with 60 natural Seeker--Assistant dialogues in contrast to synthetic or model-generated interactions; (iii) \textbf{domain specificity}, focusing on fashion, where subjective, context-sensitive choices make CRS evaluation especially challenging; and (iv) \textbf{explicit user profiles}, derived from pre-task surveys capturing style preferences, values, and behaviors. 

Our stage analysis shows a consistent five-step rhythm: Assistant-led elicitation, Seeker critique, and convergence on choices. Visual grounding and grouped recommendations support comparative reasoning and faster convergence, while transparent justification remains crucial when alignment falters.  At the item level, human Assistants still outperform large models: while MLLMs can follow explicit cues, they fail to capture relative preferences or generalize across all items. This underscores preference modeling and rating calibration as central open challenges.  

Taken together, \vogue foregrounds item-level alignment, conversational stage dynamics, and satisfaction in a realistic, visually grounded setting.  Most importantly for the development of future multimodal CRS, our findings establish \vogue as a challenge dataset beyond the current recommendation capabilities of existing top-tier MLLMs such as \textsc{GPT-4o-mini}, \textsc{GPT-5-mini}, and \textsc{Gemini-2.5-Flash}.
\begin{acks}

This paper includes limited use of generative AI tools in the data preparation and annotation pipeline. Specifically, \textsc{GPT-4o-mini} was used to summarize product descriptions and review text for catalogue items presented in the study interface. In addition, \textsc{Gemini-2.5-Flash} was used in a first-pass labeling stage to assist with large-scale dialogue intent tagging. Tags were subsequently manually reviewed and corrected by human annotators. Sec. \ref{sec:rq3} evaluates \textsc{GPT-5-mini, GPT-4o-mini, Gemini-2.5-Flash}, and \textsc{Gemini-2.0-Flash} as part of the post-conversation item-rating prediction experiments.
\end{acks}




\bibliographystyle{ACM-Reference-Format}
\bibliography{sample-base}

\end{document}